\documentclass[
twocolumn,secnumarabic,amssymb, amsmath, tightenlines,nobibnotes, aps,pre,longbibliography,superscriptaddress, floatfix]
{revtex4}

\let\oldmarginpar\marginpar
\renewcommand\marginpar[1]{\-\oldmarginpar[\raggedleft\footnotesize #1]%
{\raggedright\footnotesize #1}}

\usepackage[dvipsnames]{xcolor}   
  
\usepackage[normalem]{ulem}		

\usepackage{bm}		
\usepackage{amssymb,amsmath}
\usepackage{graphicx}
\usepackage{comment}

\usepackage{epstopdf} 
\usepackage{physics}		

\usepackage{hyperref}

\usepackage{lipsum}
\usepackage{float}

\begin{document}
\title{Optimal finite-time bit erasure under full control}

\author{Karel Proesmans}
\email[email: ]{Karel\_Proesmans@sfu.ca}
\affiliation{Department of Physics, Simon Fraser University, Burnaby, B.C., V5A 1S6, Canada}
\affiliation{Hasselt University, B-3590 Diepenbeek, Belgium}
\author{Jannik Ehrich}
\affiliation{Department of Physics, Simon Fraser University, Burnaby, B.C., V5A 1S6, Canada}
\author{John Bechhoefer}
\affiliation{Department of Physics, Simon Fraser University, Burnaby, B.C., V5A 1S6, Canada}

\begin{abstract}
We study the finite-time erasure of a one-bit memory consisting of a one-dimensional double-well potential, with each well encoding a memory macrostate.  We focus on setups that provide full control over the form of the potential-energy landscape and derive protocols that minimize the average work needed to erase the bit over a fixed amount of time.  We allow for cases where only some of the information encoded in the bit is erased.  For systems required to end up in a local equilibrium state, we calculate the minimum amount of work needed to erase a bit explicitly, in terms of the equilibrium Boltzmann distribution corresponding to the system's initial potential.   The minimum work is inversely proportional to the duration of the protocol.  The erasure cost may be further reduced by relaxing the requirement for a local-equilibrium final state and allowing for any final distribution compatible with constraints on the probability to be in each memory macrostate.  We also derive upper and lower bounds on the erasure cost.
\end{abstract}

\maketitle

\section{Introduction}
In his seminal 1961 paper, Landauer showed that a finite amount of work, known as the \textit{Landauer bound}, is needed to erase a bit of information~\cite{landauer1961irreversibility,Parrondo2015_infoTD}. This bound has been verified in several theoretical models \cite{dillenschneider2009memory,lambson2011exploring,diana2013finite} and experimental tests \cite{berut2012experimental,berut2013detailed,jun2014high,berut2015information,gavrilov2016erasure,gavrilov2017direct}. However, the Landauer bound can be saturated only in the quasi-static limit, an assumption that is never valid for practical systems.

How much extra work, then, is needed to erase a bit in a finite amount of time?  Several theoretical studies have addressed this question using assumptions such as slow driving \cite{zulkowski2014optimal,zulkowski2015optimal}, a final distribution that is locally in equilibrium \cite{aurell2012refined}, or counterdiabatic driving \cite{boyd2018shortcuts}. 

In an accompanying paper \cite{proesmans20PRL}, we have derived a general theory to optimize the work needed to erase a bit over a finite amount of time, provided that one has full control over the external potential. Often however, one is interested in  where only part of the information encoded in the bit is erased (\textit{partial erasure}) \cite{diana2013finite,zulkowski2014optimal,zulkowski2015optimal,boyd2018shortcuts,sheng2019thermodynamics}.  In this paper, we generalize the results from \cite{proesmans20PRL} to partial erasure, provide full details of the calculations, and derive a new set of upper and lower bounds associated with the finite-time erasure of a bit.

In Sec.~\ref{sec_setup}, we introduce the physical model of a bit and review the existing literature on optimal transport theory. These results are used in Sec.~\ref{secleq} to calculate the amount of work needed to erase a bit, provided that the final state is in local equilibrium. Subsequently, in Sec.~\ref{sec_nleq}, we relax the assumption on the final distribution. Our results are tested on two simple toy models in Sec.~\ref{sec_examples}. Finally, we give a thorough discussion and outlook in Sec.~\ref{sec_disc}.







\section{Setup: Optimal finite time bit erasure} 
\label{sec_setup}

Let us consider the work required to transform a system whose state $x$ is described by a probability density function $p_0(x)$ at an initial time $0$ to a density $p_\tau(x)$ at a final time $\tau$.  At the initial and final times, the potential-energy landscape is $V_0(x)$.  At intermediate times $t$, the potential may be altered arbitrarily, as specified by the function $V(x,t)$.  Jump discontinuities in space and time for $V$ are allowed.

Throughout this paper, we will assume that the dynamics of the system can be described by an overdamped Fokker-Planck equation,
\begin{equation}
    \pdv{p(x,t)}{t} = \pdv{x} \left(p(x,t)\pdv{x}V(x,t)\right) + \pdv[2]{x}p(x,t) \,,
\label{fpe}
\end{equation}
where we have scaled entropy by the Boltzmann constant, $k_\textrm{B}$, energies by the thermal energy of the environment, $k_\textrm{B}T$, lengths (squared) by Var$(x)$, the variance of $p_0(x) \equiv p(x,0)$, and time by Var$(x)/D$.  We denote the cumulative distributions associated with $p_0 (x)$ and $p_\tau (x) \equiv p(x,\tau)$ by 
\begin{align}
    f_{0/\tau}(x) = \int_{-\infty}^x \dd{x'} \, p_{0/\tau}(x') \,.
\label{eq:cdf}
\end{align} 

Our goal is to find a protocol $V(x,t)$ that minimizes the average work done on the system as it is transformed from $p_0(x)$ to $p_\tau(x)$.  In this work, we assume that the system is in global thermodynamic equilibrium initially ($t=0$).  Thus, $p_0(x)\sim\exp(-\beta V_0(x))$, where $\beta=(k_\textrm{B}T)^{-1}$ is the inverse temperature.

The analysis of Refs.~\cite{aurell2011optimal,aurell2012refined,zhang2019work, zhang2020optimization} shows that the optimal control protocol leads to the following intermediate cumulative distribution (using notation that follows Refs.~\cite{zhang2019work,zhang2020optimization}):
\begin{align} \label{eqn_ftFromf0}
    f(x,t) = f_0\left(\Gamma_t^{-1}(x)\right) \,,
\end{align}
where 
\begin{align}
    \Gamma_t(x) \equiv x + \frac{t}{\tau} 
        \left[ f_\tau^{-1}(f_0(x)) - x \right] \,.
\label{eq:Gamma-tau}
\end{align}
Applying the coordinate transformation in Eq.~\eqref{eq:Gamma-tau} leads to the intermediate probability distribution
\begin{align} 
    p(x,t) = \frac{p_0(\Gamma_t^{-1}(x))}{\Gamma_t'\left( \Gamma_t^{-1}(x) \right)} \,.
\label{eqn_intermediateProbability}
\end{align}

The potential, or protocol, $V(x,t)$ that achieves this intermediate distribution $p(x,t)$ is determined by inverting the Fokker-Plank equation~\eqref{fpe}:
\begin{align}
    V(x,t) = -\ln{p(x,t)} + \int_{-\infty}^x \dd{x}' 
        \frac{\int_{-\infty}^{x'} \dd{x}'' \, 
        \pdv{}{t}p(x'',t)}{p(x',t)} \,. 
\label{eqn_intermediatePotential}
\end{align}

When this optimal control potential is chosen, the minimum average work is \cite{aurell2011optimal,aurell2012refined,zhang2019work,zhang2020optimization}
\begin{align}
    W_{\textrm{min} \,|\, p_\tau} 
    &= \Delta \mathcal{F}+\Delta_\textrm{i}S \nonumber \\
    &=\int_{-\infty}^{\infty} \dd{x} \,
    p_\tau(x) \ln \frac{p_\tau(x)}{p_0(x)} \nonumber\\
    &\hspace{1cm} + \frac{1}{\tau}\int_0^{1}\dd{y} \,
    \left[ f^{-1}_0(y)-f^{-1}_\tau(y) \right]^2 \,,
\label{Swminmicro}
\end{align}
where the first term, $\Delta \mathcal{F}$, is the change in nonequilibrium free energy~\cite{Esposito2011NonequilibriumFreeEnergy} and the second term, $\Delta_\textrm{i} S$, the average entropy production~\cite{vandenbroeck2010ThreeFacesFokkerPlanck,seifert2012stochastic}.  Note that the cumulative probability $y = f_0(x)$ is dimensionless.  

In Eq.~\eqref{Swminmicro}, we denote the average work by $W_{\textrm{min} \,|\, p_\tau}$ to emphasize that the choice is conditioned on $p_\tau(x)$.  It also depends on $p_0(x)$, but we always fix $p_0(x)$ to be the equilibrium distribution for the given potential, $V_0(x)$. 

We now apply this general framework to the problem of bit erasure, by focusing on a bit described by a microscopic variable $x$, where the bit is in macrostate $1$ if $x>0$ and macrostate $0$ if $x<0$. We assume that the bit is initially in equilibrium with a symmetric potential $V_0(-x)=V_0(x)$, with an equilibrium distribution $p_0(x)$.  The bit thus has probability $1/2$ to be either in macrostate $0$ or $1$. 

We allow for the possibility of \textit{partial erasure}, where a fraction $\epsilon \leq 1/2$ of probability, the \textit{erasure error}, remains in the right well at the end of the protocol.  Thus, at $t=\tau$, the system is in macrostate $0$ with probability $p_\tau(x<0)=1-\epsilon$ and in macrostate $1$ with probability $p_\tau(x>0)=\epsilon$.

\section{Erasure to local equilibrium}
\label{secleq}

Previous studies of the erasure of a one-bit memory have focused on cases where, at the end of the protocol, the distribution is locally in equilibrium.  In experimental realizations, this condition arises because the manipulations are slow~\cite{berut2012experimental,berut2013detailed,berut2015information,jun2014high,gavrilov2017direct}. The assumed form of the final distribution also simplifies theoretical calculations~\cite{aurell2012refined,zulkowski2014optimal,boyd2018shortcuts}. 

In this section, we generalize such calculations to the case of partial erasure, for which
\begin{align}
    p_{\textrm{leq}}(x) = 
    \begin{cases}
        2(1-\epsilon )\,p_0(x) & x <0 \\
         2 \epsilon \, p_0(x) & x \geq 0 \,.
    \end{cases}
\label{eqn_localEQDist}
\end{align}

From Eq.~\eqref{Swminmicro}, the minimum work needed to erase a symmetric potential in finite time with a final distribution equal to the local equilibrium distribution is
\begin{align}
      W_{\textrm{min,leq}} &=W_{\textrm{min}|p_\tau =p_{\textrm{leq}}} \nonumber\\
      &= \Delta \mathcal{F}_{\textrm{leq}}+\Delta_\textrm{i}S_{\textrm{min,leq}} \,, 
\label{eqn_localEqW}
\end{align}
where $\Delta \mathcal{F}_{\textrm{leq}}$ can be evaluated using Eq.~\eqref{eqn_localEQDist}.  Taking into account the requirements that $p(x,0)=p_0(x)$ be the equilibrium Boltzmann distribution and that the potential at the end has the same form as at the beginning, $V(x,0)=V(x,\tau)=V_0(x)$, we have
\begin{align}
    \Delta \mathcal{F}_{\textrm{leq}} &= \int_{-\infty}^{\infty}\dd{x} \,    p_{\textrm{leq}}\ln\frac{p_{\textrm{leq}}(x)}{p_0(x)} \nonumber\\
    &= \ln 2+\epsilon\ln\epsilon+(1-\epsilon)\ln(1-\epsilon) \,.
\label{eqn_freeEnergyLeq}
\end{align}
Similarly, the entropy production is 
\begin{align}
    \Delta_\textrm{i}S_{\textrm{min,leq}} 
        = \frac{1}{\tau} \int_0^1 \dd{y} \, 
        \left[ f^{-1}_0(y)-f^{-1}_{\textrm{leq}}(y) \right]^2 \,,
\label{eqn_dissLocalEq}
\end{align}
where $f^{-1}_{\textrm{leq}}(y)$ is the inverse of the cumulative local equilibrium distribution $f_{\textrm{leq}}(x)$.

As we show in Appendix~\ref{app_DissLE}, Eq.~\eqref{eqn_localEQDist} implies that the entropy production can be rewritten purely in terms of the initial equilibrium distribution:
\begin{align}
    \Delta_\textrm{i}S_{\textrm{min,leq}} 
        = \frac{2}{\tau} \Bigg[ 1 - \int_{0}^{1-\epsilon}\dd{y} \, f^{-1}_0(y) f^{-1}_{0}\left(\frac{y}{2(1-\epsilon)}\right)\nonumber\\
        - \int_{1-\epsilon}^{1} \dd{y}\, f^{-1}_0(y)f^{-1}_{0} 
        \left( 1+\frac{y-1}{2 \epsilon } \right) \Bigg] \,.
\label{eq:local_erasure_work}
\end{align}
In the full-erasure limit ($\epsilon \rightarrow 0$), we obtain the compact expression
\begin{align}
    \Delta_\textrm{i}S_{\textrm{min,leq}} 
        = \frac{2}{\tau} \left[ 1 - \int_{0}^{1}\dd{y} \,f^{-1}_0(y) f^{-1}_{0}\left(\frac{y}{2}\right) \right] \,.
\end{align}

\subsection{Bounds on optimal local-equilibrium erasure cost}

In Appendix~\ref{app_DerivationUpperBound}, we show that the remaining integrals in Eq.~\eqref{eq:local_erasure_work} can be further evaluated, yielding an upper bound on the optimal erasure cost that depends only on the variance of the initial distribution:
\begin{equation}
    \Delta_\textrm{i} S_{\textrm{min,leq}}\leq 2(1-2\epsilon)\, 
        \frac{\textrm{Var}(x)}{\tau} \,.
\label{eq:supbound}
\end{equation}

Moreover, Dechant and Sakurai~\cite{dechant2019Wasserstein} recently showed that the entropy production incurred when transforming a system from an initial state $p_0(x)$ to a final state $p_\tau(x)$ over a period $\tau$ is bounded by a quantity that only depends on the first two moments of the initial and final probability distributions:
\begin{align}
    \tau\,\Delta_\textrm{i} S_{\textrm{min,leq}}  
    &\geq \langle x^2\rangle_0+\langle      
        x^2\rangle_\tau-2\left\langle x\right\rangle_0\left\langle x\right\rangle_\tau \nonumber \\
    &\quad-2\sqrt{\left(\langle x^2\rangle_0-\langle 
        x\rangle^2_0\right)(\langle x^2\rangle_\tau-\langle x\rangle_\tau^2)} \,.
\label{eq:dechant-sakurai}
\end{align}
Here, the subscripts $0$ and $\tau$ indicate averages over $p_0(x)$ and $p_{\tau}(x)$, respectively.  In the context of bit erasure, we can simplify this expression. First, because of the symmetry of $V_0(x)$, $\left\langle x\right\rangle_0=0$. Second, local equilibrium implies that $\left\langle x^2\right\rangle_\tau=\left\langle x^2\right\rangle_0=\textrm{Var}(x)$ (cf.~Eq.~\eqref{eqn_variance}). Finally, from Eq.~\eqref{eqn_localEQDist}, $\left\langle x\right\rangle_\tau=(2\epsilon - 1)\langle|x|\rangle_0$. Equation~\eqref{eq:dechant-sakurai} then simplifies to
\begin{equation}
     \Delta_\textrm{i} S_{\textrm{min,leq}} \geq \frac{2\textrm{Var}(x)}{\tau} \left(1-\sqrt{1-\frac{(1-2\epsilon)^2\langle|x|\rangle_0^2}{\textrm{Var}(x)}}\right) \,,
\label{wasbound}
\end{equation}
which leads to a lower bound for $W_{\textrm{min,leq}}$, again in terms of the initial microscopic distribution. Below, we derive a different lower bound that is valid for the work needed in optimal erasure when the final state is \textit{not} constrained to be in local equilibrium.

\section{Optimal erasure beyond local equilibrium}
\label{sec_nleq}

We now relax the assumption of a local-equilibrium final state and minimize over all final densities $p_\tau(x)$ compatible with the desired macrostate.  In contrast to the local equilibrium protocols from section \ref{secleq}, the optimal protocols in this section cannot easily be written in terms of the cumulative distribution functions. Therefore, following Refs.~\cite{aurell2011optimal,aurell2012refined,zhang2019work,zhang2020optimization}, we transform the integrals in Eq.~\eqref{Swminmicro}, which contain quantities evaluated at both the beginning and end of the protocol ($t=0$ and $\tau$) into quantities evaluated solely at time 0.  To change variables carefully, we first refine our notation so that $x_0$ is the $x$-coordinate used to describe quantities at time $t=0$, while $x_\tau$ is used for quantities at time $\tau$.  Further, we define a mapping between them [see Eq.~\eqref{eq:Gamma-tau}],
\begin{align}
    x_\tau = \Gamma_\tau(x_0) \equiv \Gamma(x_0) \,,
\end{align} 
so that the cumulative distributions obey $f_\tau(x_\tau) = f_0(x_0)$. Differentiation gives
\begin{align}
    p_\tau(x_\tau) \dd{x_\tau} &= p_0(x_0) \dd{x_0} \,,
        \nonumber \\
    p_\tau(x_\tau) &=        
        \frac{p_0(x_0)}{\left|\dv{x_\tau}{x_0}\right|}
        = \frac{p_0(x_0)}{\Gamma'(x_0)} \,.
\end{align}
Evaluating the ratio $p_\tau(x_\tau)/p_0(x_\tau)$ that appears in Eq.~\eqref{Swminmicro} in terms of $x_0$ gives
\begin{equation}
    \frac{p_\tau(x_\tau)}{p_0(x_\tau)} 
    = \frac{p_0(x_0)}{\Gamma'(x_0) p_0(\Gamma(x_0))} \,,
\end{equation}
so that
\begin{multline}
    \int_{-\infty}^{\infty}\dd{x_\tau}\, 
        p_\tau(x_\tau)\ln\frac{p_\tau(x_\tau)}{p_0(x_\tau)} 
    \\= \int_{-\infty}^{\infty} \dd{x_0} \,  p_0(x_0)\ln\left(\frac{p_0(x_0)}{\Gamma'(x_0)p_0(\Gamma(x_0))} \right) \,.
\label{Swminmicro1}
\end{multline}
With $y=f_0(x_0) = f_\tau(\Gamma(x_0))$ and $\dd{y} = p_0(x_0) \dd{x_0}$, the second integral in Eq.~\eqref{Swminmicro} becomes
\begin{align}
    \tau\,\Delta_\textrm{i} S &= \int^{1}_{0}\dd{y} \, \left[ f^{-1}_0(y)-f^{-1}_\tau(y) \right]^2 \nonumber\\
    &= \int^{\infty}_{-\infty}\dd{x_0} \, 
    p_0(x_0)\left[ x_0-\Gamma(x_0) \right]^2 \,.
\label{Swminmicro2}
\end{align}

To optimize the entropy production with respect to the final distribution, we substitute Eqs.~\eqref{Swminmicro1} and \eqref{Swminmicro2} into Eq.~\eqref{Swminmicro}.  Simplifying the notation from $x_0 \to x$, we have
\begin{multline}
    W_\textrm{min} = \min_{\Gamma(x)}
    \int_{-\infty}^{\infty} \dd{x} \, p_0(x)\times\\
    \left[ \ln \frac{p_0(x)}{\Gamma'(x)p_0(\Gamma(x))}
    +\frac{\left[ \Gamma(x)-x \right]^2}{\tau} \right] \,,
\label{Sgamw}
\end{multline}
where $W_\textrm{min}$ is the minimum average work required when $p_\tau(x)$ may vary.  In particular, we specify a class of allowed final distributions where the bit is erased with probability $1-\epsilon$,
\begin{equation}
     f_\tau(0)=1-\epsilon \,,
\end{equation}
In terms of $\Gamma(x)$ this condition can be written as $\Gamma\left( f_0^{-1}(1-\epsilon) \right) = 0$, or
\begin{equation}
    \int^{\infty}_{-\infty}\dd{x} \,\,
        \Gamma(x)\delta \left( x-f_0^{-1}(1-\epsilon) \right)=0 \,.
\label{constraint}
\end{equation}

From Appendix~\ref{appft}, this minimization leads to
\begin{multline}
    p_0(\Gamma(x))\dv{x} \left[ \frac{p_0(x)}{\Gamma'(x)p_0(\Gamma(x))} \right]
        +\frac{2p_0(x)}{\tau} \left[ \Gamma(x)-x \right] 
   \\ = \tilde{\lambda}\delta(x-f_0^{-1}(1-\epsilon)) \,.
\label{eq:p_var}
\end{multline}
If $p_0(x)$ is the Boltzmann distribution corresponding to the potential $V_0(x)$, we can rewrite Eq.~\eqref{eq:p_var} as
\begin{multline}
    V_0'(\Gamma(x)) - \frac{V_0'(x)}{\Gamma'(x)} - \frac{\Gamma''(x)}{\Gamma'(x)^2} + \frac{2}{\tau}\left[ \Gamma(x)-x \right] \\
   = \lambda\delta(x-f_0^{-1}(1-\epsilon)) \,.
\label{ts}
\end{multline}
Here $\lambda$ is a parameter that is fixed by the boundary conditions, Eq.~\eqref{constraint}, and $\Gamma(x)=x$, when $x=\pm\infty$.

\subsection{Lower bound on optimal erasure cost}

Minimizing the work, Eq.~\eqref{Swminmicro}, over all possible final distributions $p_\tau(x)$ requires a trade-off between minimizing the free energy difference (the Kullback-Leibler distance between initial and final distributions) and the entropy production.  Minimizing the former leads to the local equilibrium distribution in Eq.~\eqref{eqn_localEQDist}, which in turn implies the free energy difference given in Eq.~\eqref{eqn_freeEnergyLeq}.

\begin{figure}[ht]
\includegraphics[width=\linewidth]{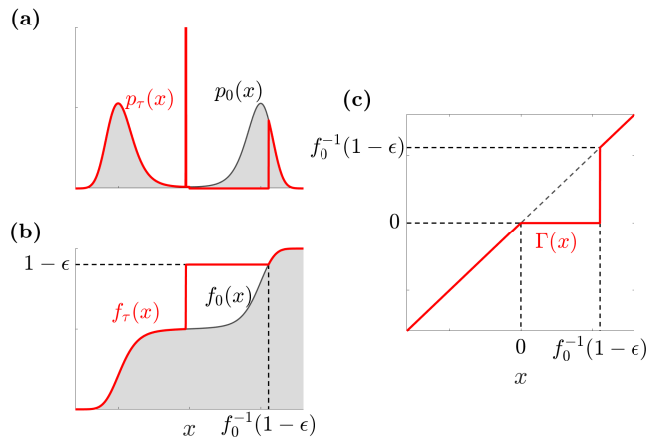}
\caption{Illustration of the final distribution compatible with the final macrostate that leads to minimum entropy production.  (a) Probability density functions $p_0(x)$ (red) and $p_\tau(x)$ (dark gray).  (b)  Cumulative distributions $f_0(x)$ (red) and $f_\tau(x)$ (dark gray).  (c) The function $\Gamma(x)$ (red) shows the coordinate transformation from time 0 to $\tau$. The curves are calculated for a quartic well [see Eq.~\eqref{Vdw}] with $E_\textrm{b}=4$ and $\epsilon = 0.1$.}
\label{fig_minDiss}
\end{figure}

Borrowing from Eq.~\eqref{Sgamw}, the latter corresponds to minimizing
\begin{equation}
    \int_{-\infty}^{\infty} \dd{x} \, p_0(x)\frac{\left[ \Gamma(x)-x \right]^2}{\tau}
\label{tomin}
\end{equation}
with respect to $\Gamma(x)$, under the constraints that $\Gamma(x)$ is non-decreasing and that Eq.~\eqref{constraint} holds. The constraints correspond to setting $\Gamma(x)\leq 0$ for $x<f_0^{-1}(1-\epsilon)$ and $\Gamma(x)\geq 0$ for $x>f_0^{-1}(1-\epsilon)$.  Minimizing Eq.~\eqref{tomin} then corresponds to minimizing the distance between $x$ and $\Gamma(x)$. As a result, the optimal $\Gamma(x)$ is 
\begin{equation}
    \Gamma(x) = 
    \begin{cases}
        x, & x < 0\, \\
        0^-, & 0 \leq x \leq f_0^{-1}(1-\epsilon) \, \\
        x, & x > f_0^{-1}(1-\epsilon) \,,
    \end{cases}
\end{equation}
as shown in Fig.~\ref{fig_minDiss}.  We can write this minimization as
\begin{align}
    \min_{\Gamma(x)}\int_{-\infty}^{\infty} \dd{x} \, 
        &p_0(x)\frac{\left[ \Gamma(x)-x \right]^2}{\tau} \nonumber \\
        &= \frac{1}{\tau} \int^{f_0^{-1}(1-\epsilon)}_0 \dd{x} \, p_0(x)x^2 \,.
\label{lowbound4}
\end{align}
This amounts to ``pushing" the necessary amount of probability density ``just over" the barrier separating the two macrostates while ``leaving behind" the probability $\epsilon$ remaining after partial erasure  (Fig.~\ref{fig_minDiss}).

A lower bound for the minimum work needed is then given by combining the individual minimizations:
\begin{eqnarray}
    W_\textrm{min}&\geq& \min_{p_\tau(x)}\int_{-\infty}^{\infty}\dd{x} \,
    p_\tau(x)\ln\frac{p_\tau(x)}{p_0(x)} \nonumber\\&&+ \min_{\Gamma(x)}\int_{-\infty}^{\infty} \dd{x} \, p_0(x)\frac{\left[\Gamma(x)-x\right]^2}{\tau}\nonumber\\
      &=&\ln 2+\epsilon\ln\epsilon+(1-\epsilon)\ln(1-\epsilon)\nonumber\\&&+\frac{1}{\tau}\int^{f_0^{-1}(1-\epsilon)}_0 \dd{x} \, p_0(x)x^2 \,.
\label{lowboundfin}
\end{eqnarray}
This bound is generally saturated in the limit $\tau\rightarrow 0$ because of the overwhelming contribution of the entropy production to the total work. 

In the limit $\epsilon \rightarrow 0$ of full erasure, we find
\begin{equation}
    W_\textrm{min}\geq \ln 2+\frac{\textrm{Var}(x)}{2\tau} \,,
\end{equation}
as discussed in the accompanying paper \cite{proesmans20PRL}.

\section{Examples}
\label{sec_examples}

In this section, we apply our general analysis to two example systems, with \textit{flat-well} and \textit{quartic} potentials.

\begin{figure*}[ht]
    \includegraphics[width = 1.0 \linewidth]{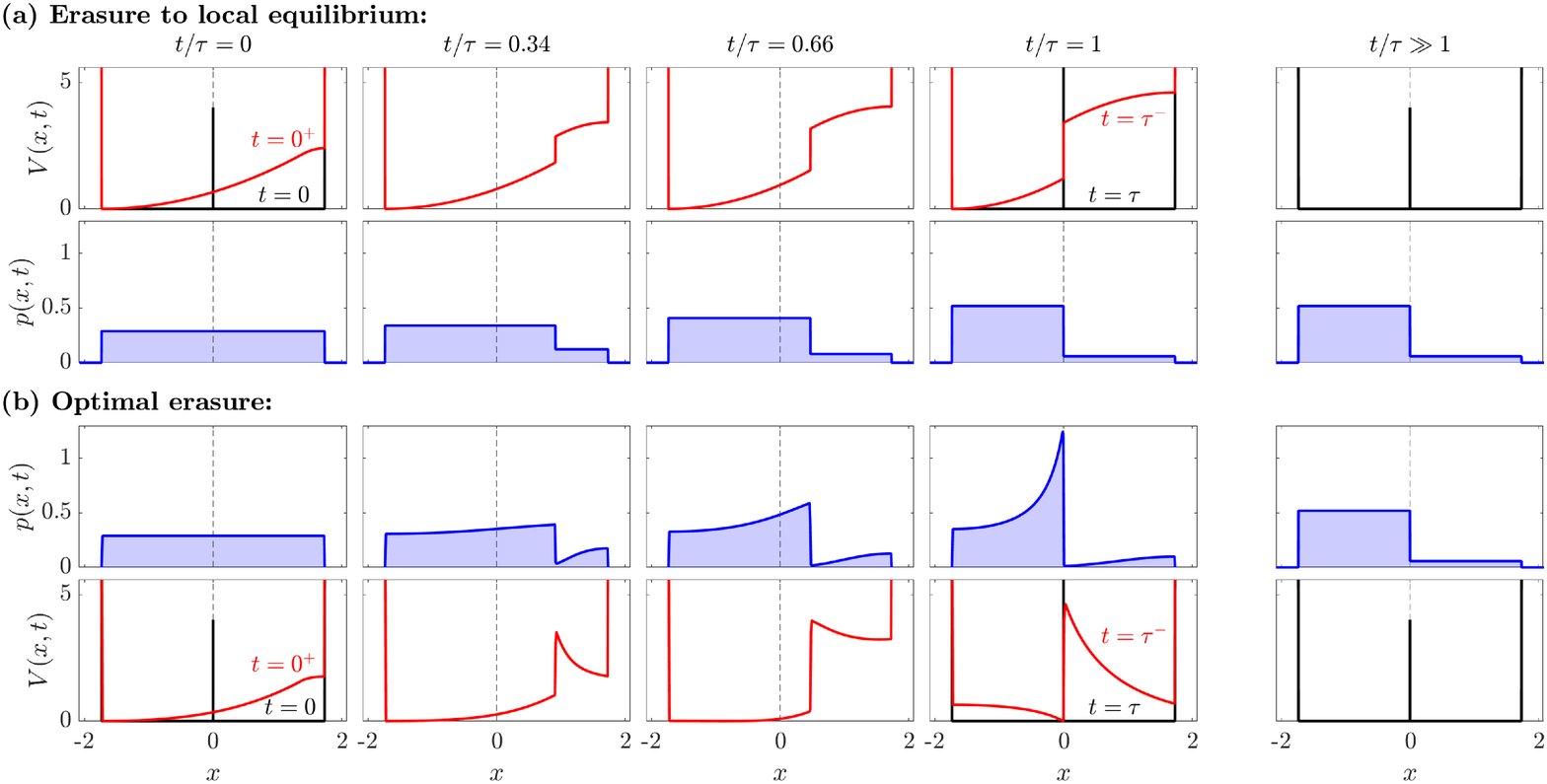}
    \caption{Partial ($\epsilon = 0.1$) finite-time ($\tau=1)$ erasure of a memory consisting of two flat wells separated by a thin barrier. (a) Potential $V(x,t)$ achieving a local equilibrium final distribution. Second row: Resulting intermediate probability distribution $p(x,t)$. Third row: Intermediate probability distribution for optimal erasure. (b) Potential achieving optimum erasure. At $t=\tau$ we prevent immediate relaxation to global equilibrium by including a potential barrier at $x=0$.}
\label{fig_flatWellIntermediatePotential_LEQ}
\end{figure*}

\subsection{Flat well}
\label{sec_flatWell}

Let us consider information erasure in a system consisting of a particle initially in equilibrium with a potential consisting of two flat wells, separated by a thin barrier. This particular example has been studied in Ref.~\cite{zulkowski2014optimal} under limited control and has the advantage that it can be evaluated analytically. For the width of the individual wells, we choose $\sqrt{3}$, so that $\langle x^2\rangle_0=1$.

In Appendix~\ref{app_flatWellCalcs}, we show that the erasure protocol that leads to a local equilibrium final distribution has the intermediate probability distribution, 
\begin{align} 
    p(x,t) &= 
        \begin{cases}
            \frac{\sqrt{3} (1-\epsilon)}
                {6 (1-\epsilon) - 3(1-2 \epsilon) t/\tau } & x < x_0(t) \\[3pt]
            \frac{\sqrt{3} \epsilon}{3 (1-2 \epsilon) t/\tau +6 \epsilon} & x > x_0(t) \,,
        \end{cases}
\label{eqn_flatwell_intermediate_prob}
\end{align}
where 
\begin{align}
    x_0(t) = \sqrt{3}(1- 2 \epsilon) \left(1- t/\tau\right) \,.    
\end{align}
Remarkably, the optimum intermediate probability distribution stays piecewise uniform throughout the erasure process.

We also calculate the intermediate potential by inverting the Fokker-Planck equation, Eq.~\eqref{eqn_intermediatePotential}, which leads to Eq.~\eqref{eqn_intermediatePotential_flatWell}. The potential is illustrated in Fig.~\ref{fig_flatWellIntermediatePotential_LEQ}(a) along with the intermediate probability distribution.

As detailed in Appendix~\ref{app_flatWellCalcs}, the entropy production when erasing to local equilibrium is given by
\begin{align}
    \Delta_\textrm{i} S_\textrm{min,leq} = \frac{(1-2 \epsilon)^2}{\tau} \,.
\label{eqn_flatWellDiss}
\end{align}
The quadratic dependence of the entropy production on the erasure error differs from the expressions found when the form of the potential $V(x,t)$ is constrained in parametric optimization \cite{zulkowski2014optimal} and when the density is constrained to be at local equilibrium for $0 \le t \le \tau$ in counterdiabatic driving \cite{boyd2018shortcuts}.  In our case, local equilibrium is imposed only at $t=0$ and $t=\tau$.

Equations~\eqref{eq:supbound},~\eqref{wasbound}, and~\eqref{eqn_flatWellDiss} then become the following set of inequalities:
\begin{align}
    2-2\sqrt{1-\frac{3(1-2\epsilon)^2}{4}} 
        \leq \Delta_\textrm{i} S_\textrm{min,leq} \, \tau 
    &= (1-2 \epsilon)^2 \nonumber \\ 
    &\leq 2(1-2 \epsilon) \,.
\label{boundsflatwell}
\end{align}

\begin{figure}[hb]
    \includegraphics[width = 0.8\linewidth]{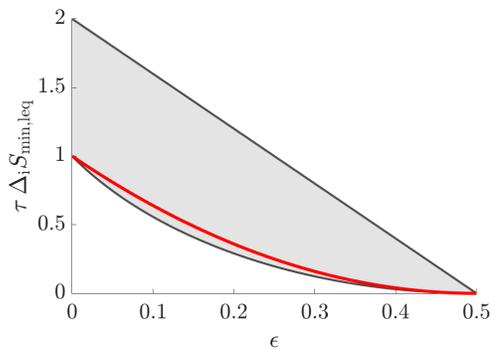}
    \caption{Comparison between entropy production (thick red curve) and its upper and lower bounds (gray curves bounding the shaded region) for two flat wells, separated by a thin barrier.}
\label{fig_LEQFlatWell}
\end{figure}

Figure~\ref{fig_LEQFlatWell} shows the entropy production together with the upper and lower bound. We immediately see that both bounds are verified and that the lower bound becomes tight in the full-erasure and zero-erasure limits. 

\subsubsection{Erasure beyond local equilibrium}

Although a flat-well potential is arguably the simplest case for bit erasure, it is still not possible to solve Eq.~\eqref{ts} analytically. Therefore, we turn to the numerical methods developed in Appendix \ref{nummet}.  The resulting optimal erasure process is illustrated in Fig.~\ref{fig_flatWellIntermediatePotential_LEQ}(b). In this protocol, a barrier emerges in the right well during the first part of the protocol. This barrier then steadily moves  to the center of the potential, pushing the probability to the left. As $\tau$ is small, the system does not have time to relax to local-equilibrium, leading to a peak in the probability distribution near the barrier. After completing the protocol, the system relaxes back to the local equilibrium distribution for $t\gg \tau$.

\begin{figure}[htb]
    \includegraphics[width = 1.0 \linewidth]{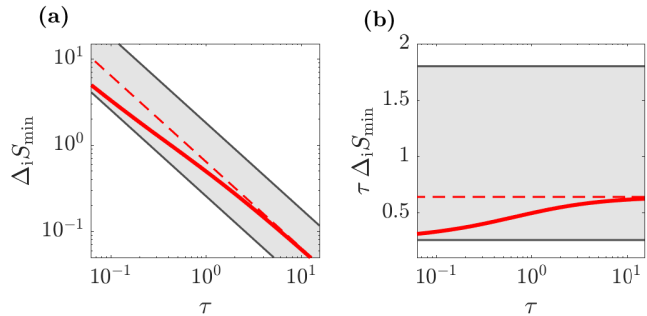}%
    \caption{Finite-time erasure costs in a flat-well potential.  (a) Comparison between $\Delta_\textrm{i} S_\textrm{min}$ (full line), $\Delta_\textrm{i} S_\textrm{min, leq}$ (dashed line) for partial erasure ($\epsilon=0.1$) of a flat well. Additionally, the upper and lower bounds, Eqs.~\eqref{eq:supbound} and \eqref{lowboundfin} (shaded region) are shown. (b) Same quantities as the left panel, but scaled by $\tau$.}
\label{fig_Wftfw}
\end{figure}

The entropy production for these various erasure protocols is shown in Fig.~\ref{fig_Wftfw}. For slow driving $W_{\textrm{min}}\approx W_{\textrm{min,leq}}$ (dashed red line).  Meanwhile, for fast driving, $\tau\rightarrow 0$, $W_{\textrm{min}}$ converges to the lower bound (lower black solid line, given by Eq.~\eqref{lowboundfin}).

For a flat-well potential, this lower bound can be evaluated analytically using Eq.~\eqref{eq:flatwell_inverse},
\begin{equation} 
    \Delta_\textrm{i} S_\textrm{min} \geq \frac{(1-2\epsilon)^3}{2} \,.
\label{eqn_minDissFlatWell}
\end{equation}
Comparing this result with Eq.~\eqref{eqn_flatWellDiss} shows that the lower bound differs from the local equilibrium result by a factor $2/(1-2\epsilon)$. This means that relaxing the assumption of local equilibrium is particularly useful in the fast driving limit.

\subsection{Quartic double well}
\begin{figure}[ht]
    \includegraphics[width=1.0\linewidth]{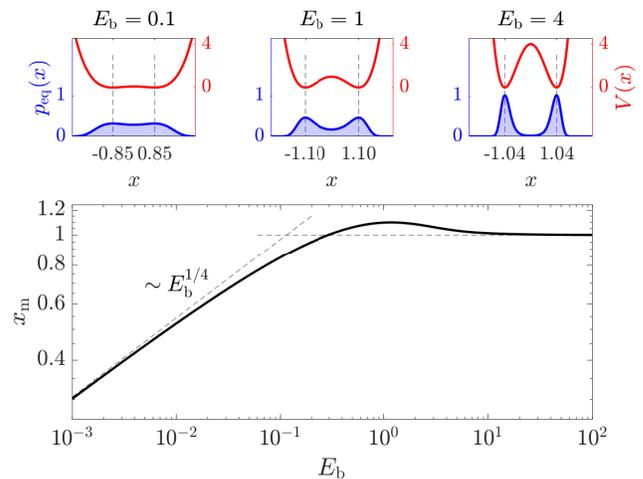}
    \caption{Equilibrium probability densities (blue shaded) for double-well potentials of energy barriers $E_\textrm{b} = 0.1, 1, 3$, in units of $k_\textrm{B}T$ (red curves).  The black curve is the value of $x_\textrm{m}$ that makes Var$_0$($x$)=1, as a function of $E_\textrm{b}$.  Dotted line indicates that $x_\textrm{m} \sim (E_\textrm{b})^{1/4}$ as $E_\textrm{b} \to 0.$}
\label{fig:xmFind}
\end{figure}

Next, consider the potential
\begin{equation}
    V_0(x) = E_\textrm{b}
        \left[ \left( \frac{x}{x_\textrm{m}} \right)^2 - 1 \right]^2 \,,
\label{Vdw}
\end{equation}
which corresponds to a double-well potential, with energy barrier $E_\textrm{b}$ and minima at $x=\pm x_\textrm{m}$ (see Fig.~\ref{fig:xmFind}). This potential has been studied extensively in the context of bit erasure, both experimentally~\cite{berut2012experimental,berut2013detailed,jun2014high,berut2015information,gavrilov2017direct} and numerically~\cite{aurell2012refined,boyd2018shortcuts}.

Here, we choose $x_\textrm{m}$ as a function of $E_\textrm{b}$ so that $\textrm{Var}(x)=1$ for all barrier heights.  The function $x_\textrm{m}(E_\textrm{b})$ is plotted in Fig.~\ref{fig:xmFind} (black curve).  For $E_\textrm{b} \ll 1$, a scaling analysis of the condition $\langle x^2 \rangle = 1$ shows that $x_\textrm{m} \approx \bigl[ \Gamma(1/4)/ (2^{1/4}\pi^{1/2})\bigr] (E_\textrm{b})^{1/4} \approx 1.72 (E_\textrm{b})^{1/4}$.  Conversely, for $E_\textrm{b} \gg 1$, the equilibrium distribution tends to two delta functions at $\pm x_\textrm{m}$, implying that $x_\textrm{m} \to 1$.

The erasure protocol $V(x,t)$ leading to a local equilibrium final distribution for $E_\textrm{b}=4$ can only be calculated numerically (see Appendix~\ref{nummet}). It is illustrated in Fig.~\ref{fig_intermediate_quartic}(a).

\begin{figure*}[ht]
    \includegraphics[width = 1.0 \linewidth]{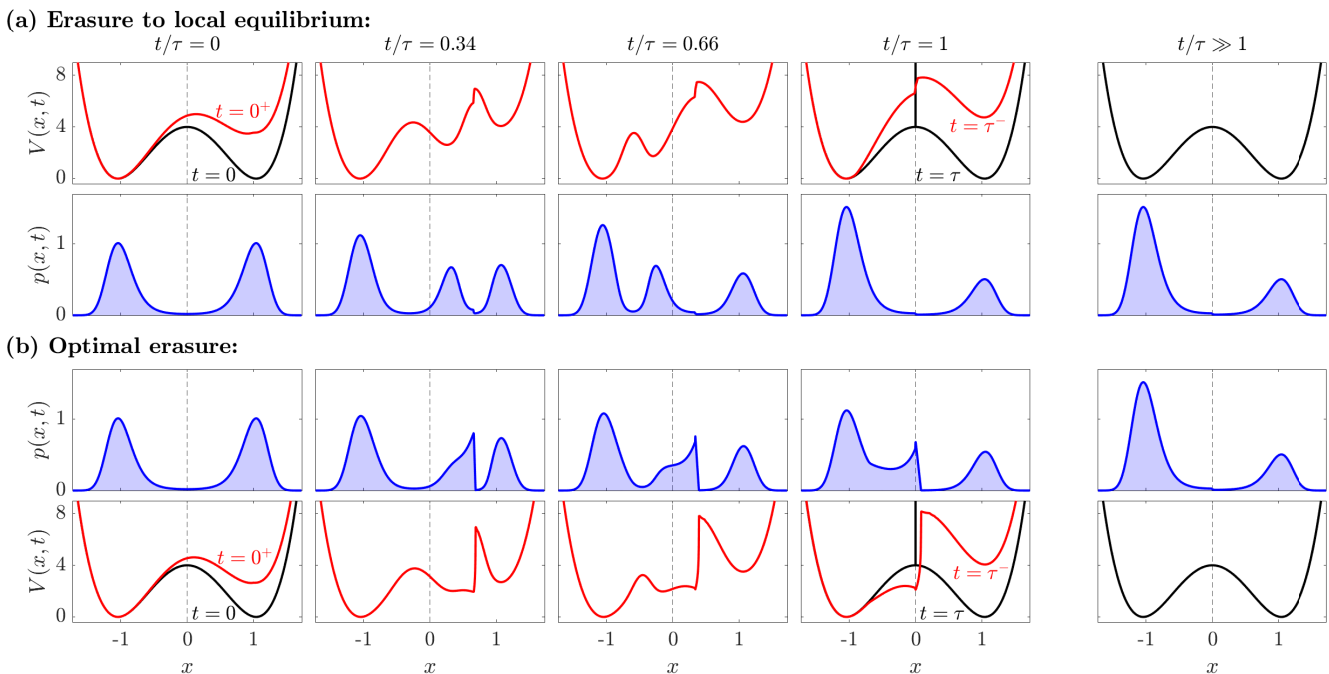}
    \caption{Partial ($\epsilon = 0.25$) finite-time ($\tau=0.5)$ erasure of a memory consisting of a quartic double well with energy barrier $E_\textrm{b}=4$. (a) Potential $V(x,t)$ achieving a local equilibrium final distribution and resulting intermediate probability distribution $p(x,t)$. (b) Intermediate probability distribution for optimal erasure and potential achieving optimum erasure. At $t=\tau$ we prevent immediate relaxation to global equilibrium by including a potential barrier at $x=0$. At $t/\tau=1$, the jump in the potential is slightly shifted to positive values due to numeric instabilities.}
\label{fig_intermediate_quartic}
\end{figure*}

\begin{figure}[hb]
    \includegraphics[width = 0.8\linewidth]{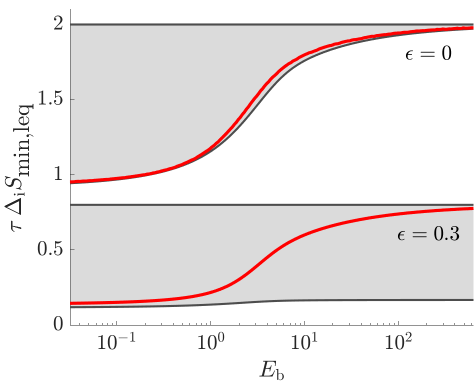}%
    \caption{Entropy production (solid lines) for partial erasure of the quartic double-well potential together with the upper and lower bounds (shaded region) as a function of the barrier height for full an partial erasure.}
\label{fig_localEQ_quartic}
\end{figure}

We numerically calculate the dissipative correction $\Delta_\textrm{i} S_\textrm{min,leq} $ to the bare Landauer cost from Eq.~\eqref{eq:local_erasure_work}. Additionally, using Eqs.~\eqref{eq:supbound}~and~\eqref{wasbound}, we numerically calculate the upper and lower bounds for the entropy production. 

The results are displayed in Fig.~\ref{fig_localEQ_quartic}. Importantly, the additional cost for optimal finite-time erasure does not scale exponentially with barrier height as claimed in Ref.~\cite{boyd2018shortcuts}.  Instead, we see that increasing the barrier height leads to a saturation of the upper bound.  This limiting case corresponds to a potential with very deep wells, which leads to an initial distribution consisting of two ``delta peaks" \cite{proesmans20PRL}.

\subsubsection{Erasure beyond local equilibrium}

\begin{figure}[ht]
    \includegraphics[width = 1.0\linewidth]{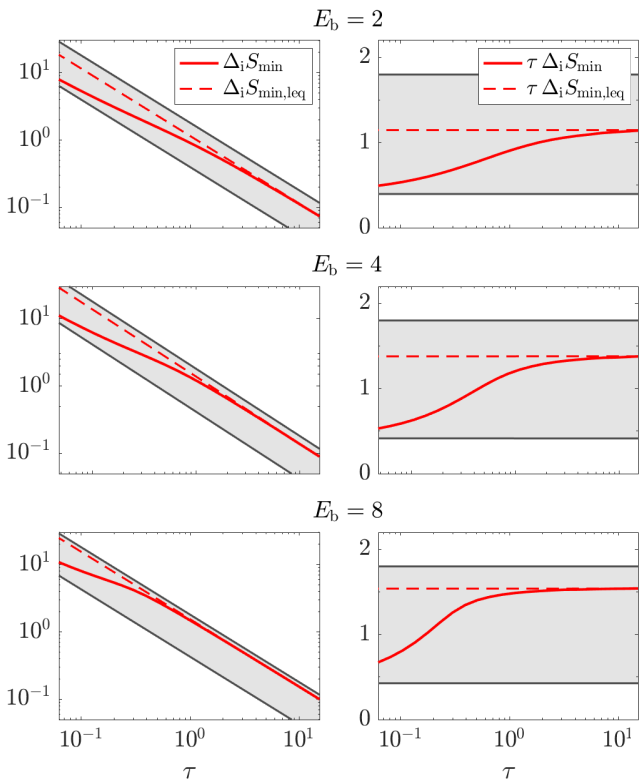}
    \caption{Finite-time erasure costs in a quartic potential.  Left panel: Comparison between $\Delta_\textrm{i} S_\textrm{min}$ (full line), $\Delta_\textrm{i} S_\textrm{min, leq}$ (dashed line) and the upper and lower bounds, Eqs.~\eqref{eq:supbound} and \eqref{lowboundfin} (shaded region) for a double well, Eq.~\eqref{Vdw} with $E_\textrm{b}=2,4,8$. Right panel: same quantities as the left panel, but scaled by $\tau$.}
\label{fig_Wftdw}
\end{figure}

Figure~\ref{fig_intermediate_quartic}(b) illustrates the optimum erasure process.  Then, in Fig.~\ref{fig_Wftdw}, we show the (numerically calculated) optimal work for finite-time erasure. One can check that, for fixed variance of the initial distribution, the erasure cost grows with a higher barrier, both in the fast and in the slow erasure limit. In the slow limit, we again find that $W_\textrm{min}$ approaches $W_{\textrm{min,leq}}$. In the fast limit, it approaches the lower bound.

\section{Discussion and Conclusion}
\label{sec_disc}


In this article, we have given a detailed treatment of the theory underlying optimization of thermodynamic processes, where the final constraint is placed at the level of a macrostate.  At the level of microstates, an entire class of distributions is allowed.  We have applied this formalism to the case of information erasure, deriving both optimal protocols and various bounds. 

Optimizing over a class of microscopic distributions can yield an optimal protocol for finite time bit erasure that uses up to a factor of four less work compared to protocols that end in local equilibrium when the bit is fully erased \cite{proesmans20PRL}.  When only part of a bit is erased, the difference in erasure cost between the local equilibrium protocol and the optimal one can be even greater, a conclusion that follows from our analysis of the flat well potential:  For short erasure processes, the minimum erasure cost tends towards the lower bound in Eq.~\eqref{eqn_minDissFlatWell}, while the local equilibrium erasure cost is given by Eq.~\eqref{eqn_flatWellDiss}.  Their ratio is a factor of $2/(1-2\epsilon)$, which diverges for $\epsilon \to 1/2$.

The method for finding the final probability distribution that results in the minimum-work erasure protocol is quite involved and can only be treated numerically even for quite simple example systems. Nonetheless, the resulting time-evolution of the probability distribution has some intuitive appeal (see Figs.~\ref{fig_flatWellIntermediatePotential_LEQ}~and~\ref{fig_LEQFlatWell}): An optimal partial erasure protocol into the left well takes the closest part of the probability distribution from the right well, pushes it over the barrier and re-distributes probability density so as to minimize a trade-off between free energy difference and entropy production.

In that context, it is noteworthy that the control potential $V(x,t)$ needed to achieve a desired evolution $p(x,t)$ of probability density is often quite involved, with steep gradients and discontinuities.  The complicated shape of the required potential leads not only to numerical difficulties (Appendix~\ref{nummet}) but also to experimental challenges.  Indeed, an important open question is whether inevitable limitations to experimental control over the potential can nonetheless allow experimental protocols whose average work closely approximate the minimum values discussed here.



One straightforward strategy to further reduce costs beyond our derived bounds would be to manipulate the intrinsic time and length scales during the protocol.  In the limit of infinitely fast relaxation times, such a strategy would enable finite-time erasure at no additional cost beyond the Landauer limit as shown, e.g., in Refs.~\cite{Browne2014,Gopalkrishnan2016}. Tempting as such an approach might be, it is not realistic.  Although reducing the length and time scales improves the performance of bit erasure, it would make little sense to have the ability to impose faster / shorter scales and only use them for part of a protocol.  Thus, we consider such a \textit{technological} optimization to have been first carried out, fixing length and time scales.  Then one proceeds to \textit{protocol} optimization, the focus of this paper.



One potential drawback of our finite-time protocols is that, for fast erasure, probability accumulates near the center of the potential and might leak back into the wrong part of the potential after completion of the protocol.  There are several ways to avoid this problem:  As already mentioned in section \ref{sec_examples}, one could add a thin, high barrier at the center of the potential immediately after completing the erasure.  The barrier would guarantee that the system relaxes to the correct macrostate and can be removed later.  Another possibility would be to push the probability slightly farther, past the center of the potential.  In this way, the system will generally relax to the correct state, with a controllable trade-off between the extra work required to push probability farther past the edge of the macrostate, against the amount of probability that will leak back into the original macrostate.  Similarly, if the goal is partial erasure, one can choose a protocol with a nominally smaller value of $\epsilon$ that will reach the desired value in the relaxation after the end of the protocol.  In this article, we have not explored such trade-offs, which all constitute sub-optimal protocols.  Our reasoning is that there are many ways to be sub-optimal, and it makes more sense to explore them in the context of a specific experimental realization, where experimental constraints, which also lead to sub-optimal strategies, can be accommodated.  In the absence of specific experimental constraints, our strategy, which achieves minimum work at the cost of needing extra time afterwards, seems a reasonable exemplar.

Finally, we note that our study here and in Ref.~\cite{proesmans20PRL} has focused on the average work to perform the single operation of bit erasure.  It would be interesting to consider the trade-offs between mean values and fluctuations~\cite{Solon2018, Maillet2019}, as well as optimizations that include a sequence of computational operations~\cite{Wolpert2019EnergeticsComputing}.


\begin{acknowledgements}
This work was supported by a Foundational Questions Institute grant, FQXi-RFP-2019-IAF, and by an NSERC Discovery Grant.
\end{acknowledgements}

\appendix

\section{Derivation of Eq.~(\ref{eq:local_erasure_work})}
\label{app_DissLE}

\begin{figure}[ht]
    \includegraphics[width = 0.9\linewidth]{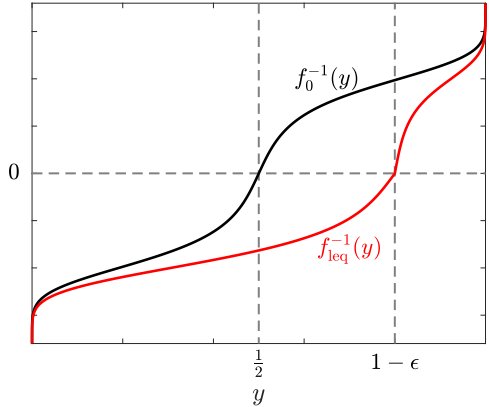}
    \caption{Example plot of the initial inverse distribution function $f_0^{-1}(y)$ and the final local-equilibrium inverse distribution function $f_\textrm{leq}^{-1}(y)$. The curves are calculated for a quartic well [see Eq.~\eqref{Vdw}] with $E_\textrm{b}=2$ and $\epsilon= 0.2$.}
\label{fig_inverseFuncsLEQ}
\end{figure}

According to Eq.~\eqref{eqn_localEQDist}, the cumulative local equilibrium distribution $f_{\textrm{leq}}(x)$, is given by
\begin{align}
    f_{\textrm{leq}}(x) &= 
    \begin{cases}
        2(1-\epsilon ) \, f_0(x) & x < 0 \\
        1-\epsilon + 2 \epsilon \int_0^x \dd{x'} p_0(x') & 
            x \geq 0
    \end{cases}\\
    &= \begin{cases}
        2(1-\epsilon )\,f_0(x) & \qquad x <0 \\
        1 + 2\epsilon\, (f_0(x)-1)  & \qquad x \geq 0.
    \end{cases} 
\label{eqn_cumulativeLocalEQ}
\end{align}
Inverting this function gives 
\begin{align}
    f^{-1}_{\textrm{leq}}(y) = 
        \begin{cases}
            f^{-1}_{0}\left(\frac{y}{2(1-\epsilon)}\right) & 0 \leq y     < 1- \epsilon \\
            f^{-1}_{0}\left(1+\frac{y-1}{2 \epsilon }\right) & 1-        \epsilon \leq y \leq 1 \,.
        \end{cases}
\label{eqn_inverseLocalEqf}
\end{align}
Figure~\ref{fig_inverseFuncsLEQ} shows an example plot of the resulting inverse functions.

Expanding the integrand in Eq.~\eqref{eqn_dissLocalEq} and inserting Eq.~\eqref{eqn_inverseLocalEqf} gives
\begin{widetext}
\begin{align}
    \Delta_\textrm{i}S_{\textrm{min,leq}} 
    = \frac{1}{\tau}\, \Bigg[ \int_{0}^{1} \dd{y}\, f^{-1}_0(y)^2 + 
    \int_{0}^{1-\epsilon} \mkern-10mu \dd{y} \, f^{-1}_0\left(\frac{y}{2(1-\epsilon)}\right)^2 +\int_{1-\epsilon}^{1} \mkern-10mu \dd{y} \, f^{-1}_0\left(1+\frac{y-1}{2 \epsilon }\right)^2 - 2\int_{0}^{1}\dd{y} \,f^{-1}_0(y)f^{-1}_{\textrm{leq}}(y) \Bigg] \,.
\end{align}
Substituting $y= f_0(x)$ in the first integral, $y/2(1-\epsilon)=f_0(x)$ in the second integral, and $1+(y-1)/2\epsilon = f_0(x)$ in the third integral yields
\begin{align}
    \Delta_\textrm{i}S_{\textrm{min,leq}} 
    &= \frac{1}{\tau}\,\Bigg[\int_{-\infty}^{\infty}\dd{x} \,p_0(x) x^2 
        + 2 (1-\epsilon) \int_{-\infty}^{0} \dd{x} \, p_0(x) x^2 
        + 2 \epsilon \int_{0}^{\infty} \dd{x} \,p_0(x) x^2 
        - 2\int_{0}^{1}\dd{y} \, f^{-1}_0(y)f^{-1}_{\textrm{leq}}(y) \Bigg] \nonumber \\
    &= \frac{2}{\tau}\, \Bigg[ \left\langle x^2 \right\rangle_0 
        - \int_{0}^{1}\dd{y} \, f^{-1}_0(y)f^{-1}_{\textrm{leq}}(y) \Bigg] \,,
\end{align}
where $\left\langle x^2 \right\rangle_0$ is the second moment of the initial distribution. Because of the symmetry of the potential (and therefore the initial distribution), it equals the variance of the initial distribution: 
\begin{align} \label{eqn_variance}
    \left\langle x^2 \right\rangle_0 = \textrm{Var}(x) \equiv 1 \,.
\end{align}

Finally, inserting Eq.~\eqref{eqn_inverseLocalEqf} for the remaining integral and recalling Eqs.~\eqref{eqn_localEqW}~and~\eqref{eqn_freeEnergyLeq}, we obtain Eq.~\eqref{eq:local_erasure_work} of the main text.

\section{Derivation of Eq.~(\ref{eq:supbound})}
\label{app_DerivationUpperBound}

We start by splitting the integrals in Eq.~\eqref{eq:local_erasure_work} further to obtain
\begin{align} \label{eqn_I12}
    \int_{0}^{1-\epsilon}\dd{y} \,f^{-1}_0(y) f^{-1}_{0}\left(\frac{y}{2(1-\epsilon)}\right) = I_1 + I_2
\end{align}
and
\begin{align} \label{eqn_I34}
    \int_{1-\epsilon}^{1} \dd{y}\, f^{-1}_0(y)f^{-1}_{0} 
        \left( 1+\frac{y-1}{2 \epsilon } \right) 
        = I_3 + I_4 \,,
\end{align}
where
\begin{align}
    I_1 &\equiv \int_{0}^{\epsilon} \dd{y} \, 
        f^{-1}_0(y) f^{-1}_{0}\left(\frac{y}{2(1-\epsilon)}\right) 
        \geq \int_{0}^{\epsilon} \dd{y} \, f^{-1}_0(y)^2 \\
    I_2 &\equiv \int_{\epsilon}^{1/2} \dd{y} \, 
        f^{-1}_0(y) f^{-1}_{0}\left(\frac{y}{2(1-\epsilon)} \right)
\end{align}
and
\begin{align}
    I_3 &\equiv \int_{1/2}^{1-\epsilon} \dd{y}\, 
        f^{-1}_0(y)f^{-1}_{0}\left(\frac{y}{2(1-\epsilon)}\right) \\
    I_4 &\equiv \int_{1-\epsilon}^{1} \dd{y}\, f^{-1}_0(y)f^{-1}_{0}
        \left(1+\frac{y-1}{2 \epsilon }\right) \geq \int_{1-\epsilon}^{1} \dd{y}\, f^{-1}_{0}
        \left(1+\frac{y-1}{2 \epsilon }\right)^2 \,.
\end{align}
The inequalities follow from the fact that $f_0^{-1}(y)$ is an increasing function and $\epsilon \leq 1/2$ (see also Fig.~\ref{fig_inverseFuncsLEQ}).

Because $p_0(x)$ is an even function, we have $p_0(-x)=p_0(x)$, and the cumulative distribution obeys
\begin{equation}
    f_0(-x)+f_0(x)=1 \,,
\end{equation}
which in turn implies
\begin{align}
    f_0(-f_0^{-1}(y))&=1-f_0(f_0^{-1}(y)) \nonumber \\
    \Longleftrightarrow f_0^{-1}(y)&=-f_0^{-1}(1-y) \,.
\label{eqn_symmetryf0}
\end{align}

Equation~\eqref{eqn_symmetryf0} can be used to rewrite $I_2$:
\begin{align}
    I_2 &= -\int_{\epsilon}^{1/2} \dd{y} \, f^{-1}_0(1-y) f^{-1}_{0}\left(\frac{y}{2(1-\epsilon)}\right) \nonumber \\
        &= -\int_{1/2}^{1-\epsilon}\dd{y'} \, f^{-1}_0(y') f^{-1}_{0} \left(\frac{1-y'}{2(1-\epsilon)}\right) \,,
\end{align}
which implies
\begin{align}
    I_2 + I_3 &= \int_{\frac{1}{2}}^{1-\epsilon}\dd{y} \, 
        f^{-1}_0(y) \Bigg[ f^{-1}_{0}
        \left(\frac{y}{2(1-\epsilon)}\right)
        - f^{-1}_{0}\left(\frac{1-y}{2(1-\epsilon)}\right)\Bigg] \,,
\label{eqn_I23}
\end{align}
Note that $f_0^{-1}(y)>0$ for $y>1/2$ and recall that $f_0^{-1}(y)$ is an increasing function. Since $y \geq 1-y$ in the interval $[1/2,1]$, it is clear that the integrand in Eq.~\eqref{eqn_I23} is positive over the entire integration domain. From this, we can deduce 
\begin{align} \label{eq:I23_positive}
  I_2 + I_3 \geq 0 \,.  
\end{align}

Next, combining $I_1$ and $I_4$ yields
\begin{align}
    I_1 + I_4 
    &\geq  \int_{0}^\epsilon \dd{y} \, f^{-1}_0(y)^2 + \int_{1-\epsilon}^{1} \dd{y}\, 
        f^{-1}_{0}\left(1+\frac{y-1}{2 \epsilon }\right)^2  \nonumber \\
    &= \int_{1-\epsilon}^{1} \dd{y'} \,f^{-1}_0(y')^2 + \int_{1-\epsilon}^{1} \dd{y}\, 
        f^{-1}_{0}\left(1+\frac{y-1}{2 \epsilon }\right)^2  \nonumber \\
    &\geq 2 \int_{1-\epsilon}^{1} \dd{y} \, 
        f^{-1}_{0}\left(1+\frac{y-1}{2 \epsilon }\right)^2 \nonumber \\
    &= 4 \epsilon \int_{1/2}^1 \dd{y'} \, f^{-1}_{0}\left(y'\right)^2 
        = 2\epsilon \left\langle x^2\right\rangle_0 \,, 
\label{eqn_lB_final}
\end{align}
where we used the symmetry relation, Eq.~\eqref{eqn_symmetryf0}, in line~2 and the fact that $f^{-1}_{0}\left(1+(y-1)/2 \epsilon)\right)$ is smaller than $f_0^{-1}(y)$ (see Fig.~\ref{fig_inverseFuncsLEQ}) in line~3.

Equations~\eqref{eq:local_erasure_work}, \eqref{eqn_variance}, \eqref{eqn_I12}, \eqref{eqn_I34}, \eqref{eq:I23_positive}, and \eqref{eqn_lB_final} together lead to
\begin{equation}
    \Delta_\textrm{i} S_{\textrm{min,leq}} \leq 2(1-2\epsilon)\, 
        \frac{\textrm{Var}(x)}{\tau} \,,
\end{equation}
which is Eq.~\eqref{eq:supbound} in the main text.

\section{Derivation of Eq.~(\ref{eq:p_var})}
\label{appft}

To minimize Eq.~\eqref{Sgamw} under the constraint Eq.~\eqref{constraint}, we define an augmented Lagrangian
\begin{align}
    \mathcal{L}
        =\int_{-\infty}^{\infty} \dd{x} \, p_0(x) \left[
        \ln \frac{p_0(x)}{\Gamma'(x)p_0(\Gamma(x))}
        + \frac{\left[ \Gamma(x)-x \right]^2}{\tau} \right]
    - \tilde\lambda \int^{\infty}_{-\infty} \dd{x} \, 
        \Gamma(x) \delta \left( x-f_0^{-1}(1-\epsilon) \right)
\end{align}
and set variations with respect to $\Gamma(x)$ equal to zero,
\begin{equation}
    \fdv{\mathcal{L}}{\Gamma(x)} = 0 \,.
\label{lagop}
\end{equation}

One can verify that, to first order in $\delta \Gamma(x)$,
\begin{align}
    \int^{\infty}_{-\infty}\dd{x} \, 
    &p_0(x) \ln \frac{p_0(x)}
        {\left[ \Gamma'(x) + \delta \Gamma'(x) \right] p_0(\Gamma(x) + \delta \Gamma(x))} 
        \nonumber \\[3pt]
    &= \int^{\infty}_{-\infty} \dd{x} \, p_0(x) \ln \frac{p_0(x)}
        {\Gamma'(x)p_0(\Gamma(x))} - \int^{\infty}_{-\infty}\dd{x} \,\, 
        \frac{p_0(x)}{\Gamma'(x)} \, \delta \Gamma'(x) 
        - \int^{\infty}_{-\infty} \dd{x} \, \frac{p_0(x)p_0'(\Gamma(x))}{p_0(\Gamma(x))}
        \, \delta \Gamma(x) \nonumber \\[3pt] 
    &= \int^{\infty}_{-\infty} \dd{x} \, p_0(x) \ln \frac{p_0(x)}
        {\Gamma'(x)p_0(\Gamma(x))} 
        + \int^{\infty}_{-\infty} \dd{x} \,\delta \Gamma(x)\, p_0(\Gamma(x)) \dv{x} 
        \frac{p_0(x)}{\Gamma'(x)p_0(\Gamma(x))} \,,
\end{align}
where we integrated by parts in the second integral of the second line to arrive at the last line. Furthermore, again to first order in $\delta \Gamma(x)$,
\begin{align}
    \int^\infty_{-\infty}\dd{x} \, p_0(x) 
        \left[\Gamma(x)+\delta \Gamma(x)-x\right]^2 
    =\int^\infty_{-\infty}\dd{x} \,\, p_0(x)
        \left[\Gamma(x)-x\right]^2 
        +2\int^\infty_{-\infty}\dd{x} \, p_0(x)
        \left[ \Gamma(x)-x \right] \delta\Gamma(x) \nonumber
\end{align}
and
\begin{align}
    \int^{\infty}_{-\infty}\dd{x} \, 
        [\Gamma(x)+\delta\Gamma(x)] 
        \delta \left( x-f_0^{-1}(1-\epsilon) \right) 
    = \int^{\infty}_{-\infty}\dd{x} \, \Gamma(x)\delta\left(x-f_0^{-1}(1-\epsilon)\right) 
        + \int^{\infty}_{-\infty}\dd{x} \, \delta\Gamma(x)\delta\left(x-f_0^{-1}(1-\epsilon)\right) \,,
\end{align}
which leads to a variation
\begin{align}
    \delta \mathcal{L} = \int^{\infty}_{-\infty}\dd{x} \, 
         \delta \Gamma(x) \,\left[p_0(\Gamma(x))\dv{x}\frac{p_0(x)}
        {\Gamma'(x)p_0(\Gamma(x))} +\frac{2p_0(x)}{\tau}\left[\Gamma(x)-x\right]
        -\tilde\lambda\delta(x-f_0^{-1}(1-\epsilon))\right] \,.
\end{align}
The Euler-Lagrange equation, Eq.~\eqref{lagop}, now leads to Eq.~\eqref{eq:p_var} in the main text.
\end{widetext}

\section{Exact calculations for the flat well potential}
\label{app_flatWellCalcs}

Here, we present the detailed calculations for the results presented in Sec.~\ref{sec_flatWell}.

\subsection{Intermediate probability distribution and potential}
The initial cumulative distribution is given by
\begin{equation}\label{eqn_flatwellCumulative}
    f_0(x)=\frac{x+\sqrt{3}}{2\sqrt{3}} \quad \text{for} \quad |x| \leq \sqrt{3} \,. 
\end{equation}

With this and Eq.~\eqref{eqn_cumulativeLocalEQ}, we eventually find the inverse of the interpolating function, Eq.~\eqref{eq:Gamma-tau},
\begin{align}
    \Gamma_t^{-1}(x) = \begin{cases}
    \frac{\sqrt{3}(1-2 \epsilon) t/\tau + 2(1-\epsilon) x}{2-(1-2\epsilon) t/\tau - 2 \epsilon} & x < x_0(t) \\[6pt]
    \frac{\sqrt{3} (1-2\epsilon) t/\tau + 2 \epsilon x}{(1- 2\epsilon) t/\tau + 2 \epsilon} & x>x_0(t) \,,
    \end{cases}
\label{appdeq1}
\end{align}
where 
\begin{align}
    x_0(t) = \sqrt{3}(1- 2 \epsilon) \left( 1- t/\tau \right)\label{appdeq2}
\end{align}
is chosen so that $\Gamma_t^{-1}(x)$ and
\begin{align}
    f(x,t) = f_0(\Gamma_t^{-1}(x))
\end{align}
are continuous.

Differentiating the cumulative distribution yields the intermediate probability distribution in Eq.~\eqref{eqn_flatwell_intermediate_prob}:
\begin{align}
    p(x,t) &=\begin{cases}
    p_l(t) & x < x_0(t)\\
    p_r(t) & x > x_0(t)
    \end{cases}\\
    &= \begin{cases}
            \frac{\sqrt{3} (1-\epsilon)}{6 (1-\epsilon) -    3(1-2 \epsilon) t/\tau } & x < x_0(t) \\[3pt]
             \frac{\sqrt{3} \epsilon}{3 (1-2 \epsilon) t/\tau +6 \epsilon} & x > x_0(t) \,.
        \end{cases}
\end{align}
Using Eq.~\eqref{eqn_intermediatePotential}, we calculate the intermediate potential
\begin{align}
    V(x,t) &= 
        \begin{cases}
            \frac{\dot{p}_l(t)}{2p_l(t)} \left(x+\sqrt{3}\right)^2 & x < x_0(t) \\[3pt]
            V_r(x,t) & x > x_0(t)
        \end{cases} \,,
\end{align}
with
\begin{align}
    V_r(x,t) &\equiv \ln\frac{p_l(t)}{p_r(t)} + 
        \frac{\dot{p}_l(t)}{2p_l(t)} 
        \left[ x_0(t)+\sqrt{3} \right]^2 \nonumber\\
    &\quad+ \frac{x-x_0(t)}{p_r(t)} 
        \Bigg\{ \dot{p}_l(t) \left[x_0(t)+\sqrt{3}\right] \nonumber \\
    &\qquad+ \frac{\dot{p}_r(t)}{2} \big[x-x_0(t)\big] 
        - \dot{x}_0(t) \big[p_r(t)-p_l(t)\big] \Bigg\} \,. 
\label{eqn_intermediatePotential_flatWell}
\end{align}
Note that the remaining constant is chosen such that the minimum of the potential is equal to zero.

\subsection{Entropy production during erasure to local equilibrium}

Inverting Eq.~\eqref{eqn_flatwellCumulative} yields
\begin{align} \label{eq:flatwell_inverse}
    f_0^{-1}(y) = \sqrt{3} \left( 2y-1\right) \quad \text{for} \quad 0 \leq y \leq 1 \,.
\end{align}
with which Eq.~\eqref{eq:local_erasure_work} can be evaluated to find Eq.~\eqref{eqn_flatWellDiss} in the main text.

To calculate the lower bound, Eq.~\eqref{wasbound}, we need the following quantity:
\begin{align}
    \left\langle |x|\right\rangle_0 
        = \int^{\sqrt{3}}_{-\sqrt{3}} \dd{x} \, \frac{|x|}{2\sqrt{3}}=\frac{\sqrt{3}}{2} \,.
\end{align}

\section{Numerical methods for finding the optimal protocol and computing the entropy production}
\label{nummet}

To find optimal protocols and associated quantities numerically, we need to solve the variational equation, \eqref{ts}, which defines a boundary-value problem.  To simplify the numerics, we limit the domain of the solution by setting $V_0(x)\rightarrow\infty$ for $x\notin \left[-x_{\textrm{max}},x_{\textrm{max}}\right]$. Then, we calculate $\Gamma(x)$ via a shooting method \cite{press07}: we fix $\Gamma(-x_{\textrm{max}})=-x_{\textrm{max}}$, and make an initial guess for $\Gamma'(-x_{\textrm{max}})$. We integrate Eq.~\eqref{ts} using a Runge-Kutta method and check whether the condition $\Gamma(f_{0}^{-1}(1-\epsilon))=0$, cf. Eq.~\eqref{constraint}, is satisfied. If $\Gamma(\Gamma(f_{0}^{-1}(1-\epsilon)))>0$, we reduce $\Gamma'(-x_{\textrm{max}})$ and integrate again.  Similarly, if $\Gamma(\Gamma(f_{0}^{-1}(1-\epsilon)))<0$ we increase $\Gamma'(-x_{\textrm{max}})$ and repeat the integration. We iterate until $|\Gamma(x_\textrm{max})-x_\textrm{max}| < 10^{-3}$. As Eq.~\eqref{ts} allows for a discontinuity in $\Gamma'(x)$ at $x=f_{0}^{-1}(1-\epsilon))$, we can repeat the above procedure to find $\Gamma'(f_{0}^{-1}(1-\epsilon))$ that leads to $\Gamma(x_{\textrm{max}})=x_{\textrm{max}}$.

Having determined $\Gamma(x)$, the remaining tasks are to find the time-dependent probability densities $p(x,t)$, the cumulative distribution $f(x,t)$, and the time-dependent potential $V(x,t)$ that implements the control.  To find the cumulative distribution $f(x,t)$, we
use Eqs.~\eqref{eqn_ftFromf0} with \eqref{eq:Gamma-tau}, which translates to
\begin{align}
    \Gamma_t(x) \equiv x + \frac{t}{\tau} \left[ \Gamma(x) - x \right] \,.
\end{align}

Having found $f(x,t)$, we can differentiate with respect to $x$ to find the corresponding density $p(x,t)$.  To find $V(x,t)$, we use Eq.~\eqref{eqn_intermediatePotential} and change the lower integration limits to $-x_\textrm{max}$.  Finally, we compute $\Delta_\textrm{i}S_\textrm{min}$ using Eq.~\eqref{Swminmicro2}. The code (in C) can be found in~\cite{note1}.

Because the map $\Gamma(x)$ can tend towards functions with discontinuous derivatives---especially for short protocols with $\tau \ll 1$---finding the numerical solution can be challenging.  We checked the shooting method described above with the built-in Mathematica command \texttt{NDSolveValue}, which claims to internally implement the shooting method as a root-finding problem. 

In addition, we discretized $p_\tau(x)$ and wrote code to estimate $W$ using Eq.~\eqref{Swminmicro}. We then minimized $W$ by gradient descent, directly varying the discretized values of $p_\tau(x)$.  Both of these approaches produced results that agree with those presented in the paper.  However, all methods fail for small-enough $\tau$ and/or large-enough $E_\textrm{b}$, where solutions can become more singular.

\end{document}